# Improvement of critical current in MgB$_2$/Fe superconducting wires by a ferromagnetic sheath

J. Horvat, X. L. Wang, S. Soltanian, S. X. Dou

Institute for Superconducting and Electronic Materials, University of Wollongong, NSW 2522, Australia

Transport critical current ($I_c$) was measured for Fe-sheathed MgB$_2$ round wires. A critical current density of 5.3 x 10$^4$ A/cm$^2$ was obtained at 32K. Strong magnetic shielding by the iron sheath was observed, resulting in a decrease in $I_c$ by only 15% in a field of 0.6T at 32K. In addition to shielding, interaction between the iron sheath and the superconductor resulted in a constant $I_c$ between 0.2 and 0.6T. This was well beyond the maximum field for effective shielding of 0.2T. This effect can be used to substantially improve the field performance of MgB$_2$/Fe wires at fields at least 3 times higher than the range allowed by mere magnetic shielding by the iron sheath. The dependence of $I_c$ on the angle between field and current showed that the transport current does not flow straight across the wire, but meanders between the grains.

Soon after discovery of superconductivity in $MgB_2$[1], superconducting wires were produced with large macroscopic critical current densities $(J_{c0})$[2, 3, 4]. Fe sheathed $MgB_2$ wires are currently the most promising conductors, giving a $J_{c0}$ of about 10 $kA/cm^2$ above 30K in a 1T field[5]. Majoros et al. predicted theoretically that magnetic shielding could decrease transport ac loss[6]. A model by Genenko et al.[7] predicted either a suppression or enhancement of the loss-free transport current of a superconducting strip in magnetic surroundings. Earlier measurements[5, 8] indicated the existence of magnetic shielding in $MgB_2$/Fe tapes. In this paper, we present the first detailed measurements of the field dependence of transport $J_c$, influenced by magnetic shielding and interaction between the Fe sheath and the superconductor in round $MgB_2$/Fe wires. It was found that interaction between the sheath and superconductor leads to a plateau in $J_c(H)$, which can be utilised to improve $J_c$ in fields three times higher than the maximum field for which the magnetic shielding of the sheath is still effective.

Three superconducting Fe-sheathed round wires were prepared by filling the Fe tubes with a mixture of Mg and B powder, drawing them into thin wires, and heating in argon for 10 minutes at 820, 890 and 850° C for samples S1, S2 and S3, respectively. The details of the wire production are described elsewhere[5]. The dimensions of the samples and $J_{c0}$ at 32K are given in Table 1.

Voltage-current characteristics were measured using a 6 milliseconds long pulse of current. The current was swept at a constant rate, with the maximum current 250A. The signal from the voltage taps was filtered by a low-pass filter and pre-amplified by a SR560 preamplifier. The current through the wire was measured via the voltage drop on a non-inductive resistor, connected in series to the wire and current source. Both voltage and current signals were fed into a digital oscilloscope. The measured data were transferred into a computer for analysis. Using a high enough cut-off frequency of the low-pass filter prevented distortions of the voltage signal and the phase shift between the current and voltage signals.

The measured $MgB_2$/Fe wire was placed in a continuous flow cryostat, with temperature control better than 0.1K. The cryostat was placed in an electromagnet on a rotating base, enabling the angle between the field and long axis of the wire to be changed. Measurements were limited to the temperature range between 32 and 35K, due to the limitations of the current source. V(I) measurements showed a very sharp increase in the voltage at the critical current ($I_c$).

Figure 1 shows the angular dependence of $I_c$ for S1, at 33.7K and 0.4T. $I_c$ decreased by 75% of its maximum value within 30° from the perpendicular orientation ($\theta=90°$). However, only a negligible change in $I_c$ was obtained for the remaining 60° (Fig.1). These measurements helped to accurately align the field into a perpendicular orientation. The inset to Fig. 1 shows the temperature dependence of $I_c$ for S1 in zero field. $I_{c0}$ decreased linearly with temperature between 32 and 36 K, at a rate of 46.4 A/K. For higher temperatures, $I_c$ decreased more gradually, approaching zero at about 38.5K.

Figure 2 shows the field dependence of $I_c$ for sample S3 at 32K. The solid and open symbols are for perpendicular ($\theta=90°$) and parallel field, respectively. In the latter case, the field is parallel to the current. The solid line shows the value of the self-field produced by the critical current at the surface of the superconducting core. For the field oriented parallel to the wire, $I_c$ does not change with the field up to about 0.03T (open symbols in Fig.2). For higher fields, an exponential decrease in $I_c$ was obtained: $I_c = I_{c0} \exp(-H/H_0)$. For all the samples measured, $H_0 \approx 0.35T$ at 32K.

For the perpendicular field, the field dependence of $I_c$ was the same as for the parallel field for H>0.6T. The inset to Fig.2 shows that the experimental points for the two field orientations overlapped by adding 0.38T to parallel field. However, for 0.2T < H < 0.6T, there was a plateau in $I_c(H)$, where $I_c$ decreased by less than 5% of $I_{c0}$ (Figure 2). For H< 0.2T, $I_c$ decreased with field by about 20% of $I_{c0}$.

The same results were obtained for the other samples measured, except for the difference in $I_{c0}$ and the field by which $I_c(H)$ with parallel field had to be shifted to obtain overlapping with $I_c(H)$ with perpendicular field ( Inset to Fig. 2). The values of this field for S1 and S2 were 0.39T and 0.33T, respectively.

The results shown in Fig.2 are affected by the magnetic shielding due to the Fe sheath, as well as by the interaction between the sheath and superconductor. To identify the effects of shielding only, the field inside and outside the sheath was measured, with the $MgB_2$ removed. This was performed by inserting tiny pick-up coils into the sheath and using external ac magnetic field with frequencies between 20 and 60 Hz, and a field amplitude up to 0.6T. The length of the coils corresponded to the distance between the voltage contacts when measuring $I_c(H)$. Comparing the results for different frequencies, we obtained that the dynamic effects (eddy currents) were negligible below 30Hz.

Figure 3 shows the field inside the sheath against the perpendicular external field (open symbols). The solid symbols were measurements with the sheath removed from the pick-up coil ($H_{in}=H_{out}$) and the solid line shows theoretical shielding[9] for an infinite cylinder of the same dimensions and magnetic susceptibility as our Fe sheath. For H<0.2T, the shielding from external field was almost total, with $H_{in}$= 0.04 $H_{out}$ at 0.2T (Fig.3). For higher fields, the shielding rapidly weakened and for H>0.4T the entire external field additional to 0.4T was passed through the Fe sheath, i.e. $H_{in}$

against $H_{out}$ was parallel to the data with no shield for H>0.4 T. These measurements are in good quantitative agreement with calorimetric measurements of ac loss in a similar $MgB_2$/Fe wire[8]. However, the measured shielding is better than that given by the analytical expression for an infinite cylindrical shield of the same thickness and magnetic permeability[9] (solid line in Fig.3). Still, extrapolation of the experimental results to high fields is in agreement with the theoretical prediction. The discrepancy at low fields is probably due to the finite length of the measured sheath.

The inset to Fig.3 shows the measured shielding for the parallel field. The dashed line represents $H_{in}=H_{out}$. The shielding was almost total for H< 0.02T. For H>0.025T, the entire field higher than 0.025T was passed through the shield. The hysteresis was due to magnetic hysteresis of the iron.

$I_c(H)$ for the parallel field can be explained by the shielding effect. The initial plateau is a consequence of complete shielding from the external field. Above 0.025T, $I_c(H)$ is the same as with no shield, except for about 0.025T which is screened out by the shield. The same is obtained for the perpendicular field for H>0.6T, except that the value of the screened-out field is about 0.3T (Fig.3). However, $I_c(H)$ for H<0.6T cannot be explained by simple screening. Instead of the expected constant $I_c$ for the external field fully screened out (H<0.2T), $I_c$ actually decreases with the field (Fig. 2). For 0.2T<H<0.6T, $I_c$ decreases very little with field (Fig.2), despite the full penetration of field through the Fe sheath (Fig.3).

Overlapping of $I_c(H)$ for parallel and perpendicular fields above 0.6T shows that the current does not flow through the wires in a straight line. If that were the case, $I_c(H)$ would differ for the two orientations of the field because the current parallel to the field would give a Lorentz - force - free configuration. Apparently, the current meanders between the grains in the wire, resulting in the average Lorentz force being the same for the two field directions. The decrease in $I_c$ for H<0.2T cannot be ascribed to weak links, because such a decrease was not also observed for the parallel field (Fig.2).

In conclusion, we demonstrate by direct transport measurements of $I_c$ that an iron sheath can be used as a very effective magnetic shield with $MgB_2$ superconducting wires. The initial decrease in $I_c$ with field and a plateau in intermediate fields is a newly observed effect, originating in an interaction between the Fe sheath and superconductor. Better understanding of this effect can lead to extending the plateau to higher fields and improving the field performance of $MgB_2$/Fe wires further. It was also shown that the current path in the wires meanders between the grains.

| Sample | $d_o$ (mm) | $d_i$ (mm) | $l$ (mm) | $J_{c0}$ (A/cm$^2$) |
|--------|------------|------------|----------|---------------------|
| S1     | 1.50       | 0.85       | 14       | 38,700              |
| S2     | 1.52       | 0.95       | 14       | 21,000              |
| S3     | 1.30       | 0.65       | 13       | 53,300              |

Table 1: Dimensions of the samples measured: $d_o$, $d_i$, and $l$ are outer diameter, inner diameter and length, respectively. $J_{c0}$ is the critical current density in zero external field, at temperature 32K.

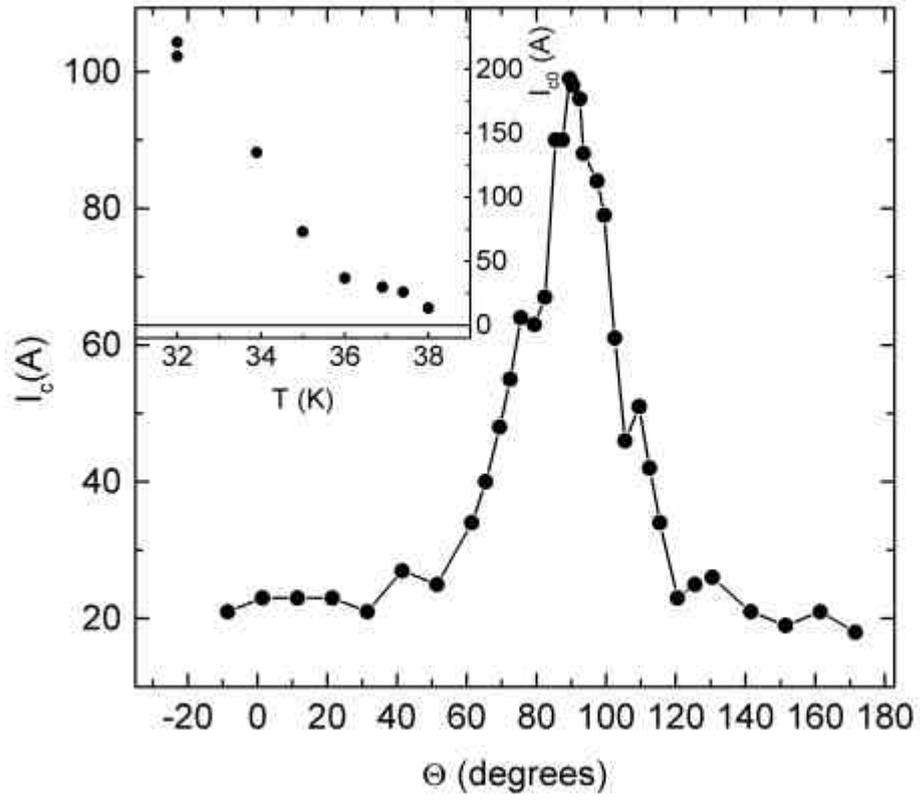

Figure 1: Angular dependence of critical current for sample S1 at 33.7K and 0.4T.

Inset: Temperature dependence of critical current in zero field for S1.

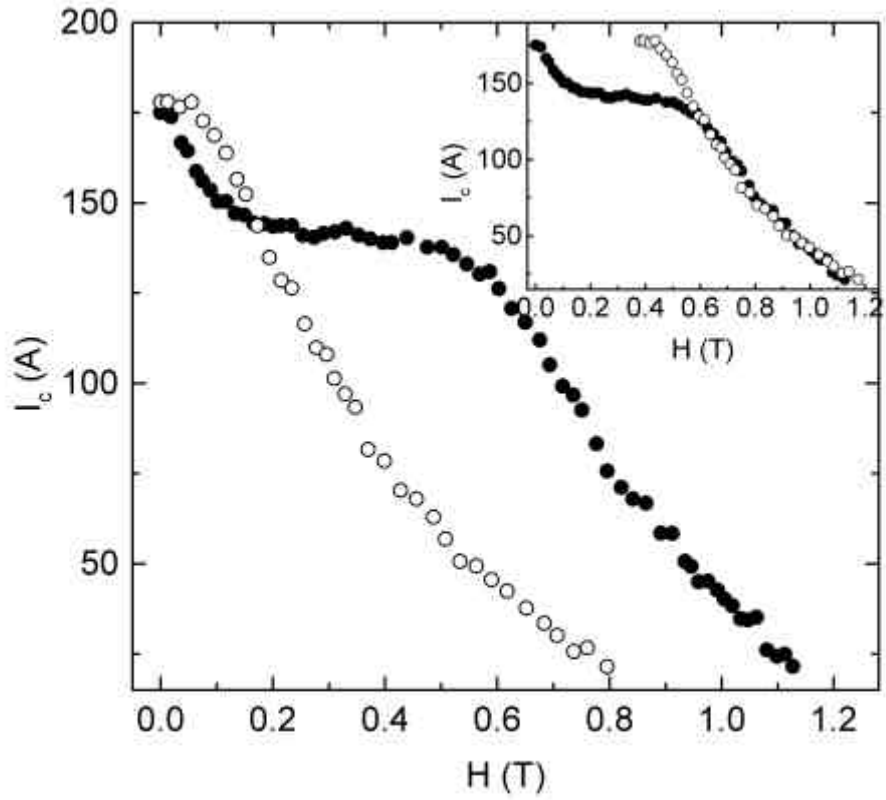

Figure 2: Field dependence of critical current for sample S3 at 32K. The solid and open symbols are for perpendicular and parallel field, respectively. The solid line is the self-field produced on the surface of the superconductor by the critical current. Inset: The same, with 380mT added to the parallel field.

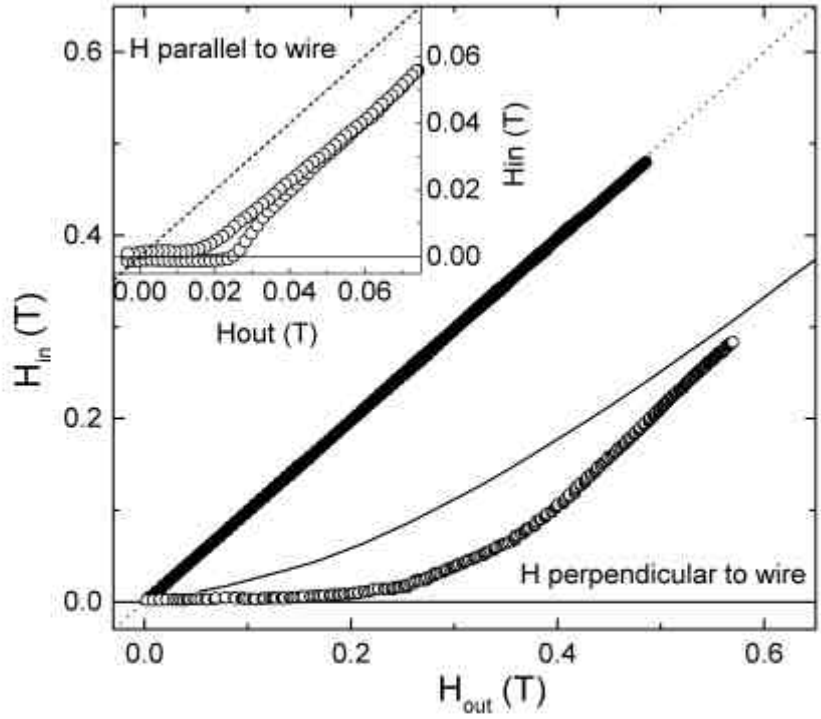

Figure 3: The magnetic field inside the iron sheath, $H_{in}$, against the external field, $H_{out}$, for perpendicular field (open symbols). When the iron sheath is removed, $H_{in}=H_{out}$ (solid symbols). The solid line shows theoretical $H_{in}$ against $H_{out}$. Inset: $H_{in}$ against $H_{out}$ for a parallel field (solid symbols). The dashed line shows $H_{in}=H_{out}$.

References:

[1] J. Nagamatsu, N. Nakagawa, T. Muranaka, Y. Zenitani, J. Akimitsu, Nature **410**, 63 (2001).

[2] P. C. Canfield, D. K. Finnemore, S. L. Bud'ko, J. E. Ostenson, G. Laperot, C. E. Cunningham, C. Petrovic, Phys. Rev. Lett. **86**, 2323 (2001).

[3] G. Grasso, A. Malagoli, C. Ferdeghini, S. Roncallo, V. Braccini, M. R. Cimberle, A. S. Sirri, Appl. Phys. Lett. **79**, 230 (2001).

[4] B. A. Glowacki, M. Majoros, M. Vickers, J. E. Evetts, Y. Shi, I. McDougall, Supercond. Sci. Technol. **14**, 193 (2001).

[5] S. Soltanian, X. L. Wang, I. Kusevic, E. Babic, A. H. Li, M. J. Qin, J. Horvat, H. K. Liu, E. W. Collings, E. Lee, M. D. Sumption, S. X. Dou, Physica C **361**, 84 (2001).

[6] M. Majoros, B. A. Glowacki, A. M. Campbell, Physica C **334**, 129 (2001)

[7] Yu. A. Genenko, A. Snezhko, H. C. Freyhardt, Phys. Rev. B **62**, 3453 (2000).

[8] M. D. Sumption, E. Lee, E. W. Collings, X. L. Wang, S. X. Dou, "Suppression of AC (hysteretic) loss by magnetic shielding of $MgB_2$/Fe superconductors: the pseudo-Meissner effect", Proc. of ICMC 2001 Conference, Madison, Wisconsin, 16-21 July 2001, in press.

[9] T. Rikitake, *Magnetic and Electromagnetic shielding* (Terra Scientific Publishing Co., Tokyo, 1987)p.43.